\documentstyle[prc,preprint,tighten,aps]{revtex}
\begin {document}
\draft
\title{
Comment on ``Large-space shell-model calculations for light
nuclei''}
\author{Attila Cs\'ot\'o$^{1,}$\cite{email} and Rezs\H o G. Lovas$^2$}
\address{$^1$National Superconducting Cyclotron Laboratory,
Michigan State University, East Lansing, Michigan 48824 \\
$^2$Institute of Nuclear Research of the Hungarian
Academy of Sciences, Debrecen, H--4001, Hungary}
\date{\today}

\maketitle

\begin{abstract}
\noindent
In a recent publication Zheng, Vary, and Barrett
reproduced the negative quadrupole moment of $^6$Li
and the low-lying positive-parity states of $^5$He by using
a no-core shell model. In this Comment we
question the meaning of these results by pointing out that
the model used is inadequate for the reproduction of these
properties.
\end{abstract}

\narrowtext

Recently, Zheng, Vary and Barrett have published a number of
no-core shell-model calculations for light
nuclei in a series of papers; see \cite{Zheng} and references
therein. In their most recent article \cite{Zheng} they presented
results for two long-standing problems of
light nuclei. Namely, they reproduced
the negative quadrupole moment of $^6$Li and found the
previously predicted low-lying positive-parity states of
$^5$He. In this Comment we argue that these results cannot be
considered the solutions to the problems in question since the model
is not realistic enough to cope with these problems. We first discuss
the quadrupole moment of $^6$Li and then the positive-parity states of
$^5$He.

The experimental value of the $^6$Li quadrupole moment is
$-0.083$ $e$$\cdot$fm$^2$ \cite{Ajzenberg}. There have been numerous
theoretical calculations attempting to reproduce this negative value,
without success. They include variational three-body calculations
\cite{Kukulin}, the hyperspherical harmonics expansion method
\cite{Danilin}, the Faddeev approach \cite{Schelling}, and a six-body
$\alpha+p+n$ three-cluster model \cite{Li6}.
All these calculations gave positive values for
$Q$, in the 0.2--0.6 $e$$\cdot$fm$^2$ range. In Ref.~\cite{H1}
it was claimed that a three-cluster model can reproduce the
negative quadrupole moment of $^6$Li if the full six-body
antisymmetrization is properly taken into account. This claim
was, however, disputed \cite{Li6}, and the result was shown to arise
from a restriction on the model space (see also
\cite{H2}). These macroscopic and microscopic $\alpha +p+n$
calculations revealed that the value of the $^6$Li quadrupole moment
results from a delicate balance between the contributions coming
from configurations of different angular momenta in the $p$--$n$
and $\alpha$--$(pn)$ relative motions \cite{Li6,Schelling}.
This indicates that, for $Q$ to be reliable, the $p$--$n$ and
$\alpha$--$(pn)$ dynamics must be described correctly.

In Ref.~\cite{Zheng} the model produces $Q=-0.116$ $e$$\cdot$fm$^2$.
It would be the first time that a realistic
microscopic model reproduced the correct negative sign of
$Q$. However, in Table I of Ref.~\cite{Zheng} we can
see that the ground-state binding energy of $^6$Li is
incompatible with those of the deuteron and of
the $\alpha$ particle. As a consequence, the model $^6$Li
seems to be unstable against the $\alpha +d$ breakup by 0.21
MeV, in sharp contrast with reality. (The
ground state of $^6$Li is below the $\alpha +d$ threshold by
1.475 MeV \cite{Ajzenberg}.) In fact, the separation energy should be
calculated from binding energies obtained in fully consistent state
spaces, which is not exactly so in Ref.~\cite{Zheng}.
The state space used for $^6$Li is restricted by
2$\hbar\omega$ more than those of $\alpha$ and $d$, thus the breakup
energy implied by the $^6$Li wave function may differ from 0.21 MeV.
Nevertheless, it looks likely that such a major discrepancy cannot
result from a such a minor mismatch between the state spaces. To
reproduce the sort of loose {\it intercluster} binding that is
characteristic of the $\alpha$--$d$ relative motion is a major challenge
to any shell model \cite{note}.

Since the $^6$Li wave function contains several configurations, each
belonging to a specific separation energy, it would be difficult to
demonstrate the strong dependence of the quadrupole moment on
all these separation energies. Just to give an indication for
the magnitude of such a dependence, in Fig.\ 1 we show the
quadrupole moment of the ground state of $^7$Li in an $\alpha +t$
cluster model, using the Minnesota effective nucleon-nucleon
force \cite{Tang}. To mimic the situation in
Ref.\ \cite{Zheng}, the exchange mixture parameter of the
interaction is varied such that the energy of $^7$Li
changes, while the binding energies of $\alpha$ and $t$ remain fixed.
We show $Q$ as a function of the $\alpha$+$t$ separation energy.
One can see that the quadrupole moment is very sensitive to this
quantity. The $^6$Li quadrupole moment is likely to behave in a
similar way, so that an increase of the $\alpha +d$ separation energy
by some 1.68 MeV could considerably shift it toward positive values.

In summary, a theoretical value for the $^6$Li quadrupole moment
can only be accepted as a physically meaningful value if
the model produces the correct separation energies,
especially for the most important $\alpha+d$ partition.
Without this, a reproduction of the experimental value can only be
fortuitous.

The problem with the low-lying positive-parity states of $^5$He in
Ref.~\cite{Zheng} is that the model cannot handle the asymptotics of
the wave function correctly.
In Ref.~\cite{Zheng} the zero point of the energy scale is at the
total $A$-body disintegration threshold of the $A$-nucleon
system. So, any state which is below this $A$-body threshold
has a negative energy, which would imply an exponentially damped
bound-state $A$-body asymptotic wave function \cite{Delves}
\begin{equation}
\Psi_A \sim \exp (-k_A\rho _A)\ \mbox{for}\
\rho_A\rightarrow\infty.
\end{equation}
Here $\rho_A$ is
the hyperradius, $\rho_A^2=\sum_ir_i^2$, where $r_i$ are the
one-particle position vectors, and $k_A=(2m_NE_A/\hbar^2)^{1/2}$,
where $E_A$ is the binding energy of the $A$-body system relative
to the $A$-body disintegration threshold, and $m_N$ is the nucleon
mass. Such a $\Psi_A$ can really be expanded in terms of
square-integrable functions, like in a shell model.
However, the Eq.\ (1) boundary condition is correct, describing
a physical situation, only if there is no break-up channel
below $E_A$. If there is a two-body ($A=B+C$) break-up channel
below $E_A$, then the correct boundary condition is
\cite{Merkuriev}
\begin{eqnarray}
\Psi_A \sim\exp (-k_A\rho _A)
+&& \Phi^B\Phi^C\left [x\exp (-ikr)+y\exp (ikr)\right]
\nonumber \\
&&\mbox{for }\rho_A,r \rightarrow \infty.
\end{eqnarray}
Here $r$ is the distance between the fragments $B$ and $C$, the
functions $\Phi$ are the internal states of $B$ and $C$ with binding
energies $E_B$ and $E_C$, and $k=[2\mu (E_A-E_B-E_C)/\hbar^2]^{1/2}$.
A scattering ``state", of energy $E_A$, that obeys Eq.\ (2) will be
regarded as a (resonant) {\it state of the nucleus} if $S=-y/x$
(the ``S-matrix") has a pole at the complex energy $E_A-i\Gamma /2$,
where $\Gamma$ is the total width.

Square-integrable bases, like that of Ref.~\cite{Zheng}, are
obviously unable to observe the boundary condition in Eq.\ (2),
so whatever they predict for states above break-up thresholds
is to be taken with reservation. This applies to the low-lying
positive-parity states of $^5$He, which are above the
$\alpha +n$ threshold.

We have performed a search for such states in a cluster
model whose wave function is
\begin{eqnarray}
\Psi &=&\sum_{S,L}\left[{\cal A}\left\{\left[(\Phi^\alpha\Phi^n)_S
\chi ^{\alpha n}_L(\mbox{\boldmath $\rho $}_{\alpha n}
)\right ]_{JM} \right \} \right.\nonumber \\
&+& \left.{\cal A}\left \{\left [(\Phi^d\Phi ^t)_S
\chi ^{dt}_L(\mbox{\boldmath $\rho $}_{dt})\right ]_{JM}
\right \}\right],
\end{eqnarray}
\noindent
where ${\cal A}$ is the intercluster antisymmetrizer,
the cluster internal states $\Phi$ are translation
invariant $0s$ harmonic oscillator shell-model states for
the $\alpha$ particle, deuteron, and triton, the vectors
\mbox{\boldmath $\rho $} are the intercluster Jacobi
coordinates, and [...] denotes angular momentum coupling.
In the sum over $S$ and $L$ all possible configurations
are included. The intercluster relative-motion functions
$\chi$ have the correct asymptotics. To avoid any
ambiguity in the recognition of a resonance in the
phase shift $\delta={1\over 2}\arg S$, we searched for
complex-energy poles of the S-matrix directly.
Both an analytic continuation method \cite{PRL} and the complex
scaling method \cite{CSM} were used. The $3/2^-$ and $1/2^-$ states
were found, but the next level was the $3/2^+$ state at
16 MeV excitation energy. No sign for any low-lying $1/2^+$,
$3/2^+$, or $5/2^+$ states was found. The inclusion of
a few monopole breathing excitations of $\alpha$ did not
change the situation either. This rules out even the exotic
possibility that the low-lying positive-parity states are
Pauli resonances \cite{Saito}, since with the departure
from the single oscillator description of the $\alpha$
cluster, the configurations that might produce Pauli
resonances get automatically included.

To produce a low-energy $1/2^+$ state artificially, we made the
intercluster binding stronger, while keeping the cluster binding
energies fixed, by changing a mixing parameter of the $N$--$N$
interaction. Then we let the mixing
parameter tend toward its physical value, and followed the
position of the $1/2^+$ pole. The pole moved rapidly towards
higher energies and, for example, it was found at 19 MeV
excitation energy with a width of 34 MeV while the mixing parameter
was still highly non-physical; indeed, the same parameter value
produced deeply bound $3/2^-$ and $1/2^-$ states. Further change
of the mixing parameter in the direction of its correct value pushed
this $1/2^+$ state to even higher energies with larger widths.

All in all, in a model that handles the asymptotics correctly, the
low-lying positive-parity states of $^5$He do not show up. Although
the basis used by Zheng {\it et al.} \cite{Zheng} is probably more
flexible than ours to describe the correlated short-range
motion of the nucleons, it is very difficult to imagine that an
improvement of our model in this direction would bring down
high-lying positive-parity states from the upper region of the
continuum. Nevertheless, it is still to be proven whether or not the
non-physical boundary condition results in the appearance of these
states. The best way to check this would be to supplement the
wave function of Ref.~\cite{Zheng} by an $\alpha +n$ cluster term
which describes the correct asymptotics. This could be done, for
example, in a cluster-configuration shell-model~\cite{VK}. An
indirect indication as to whether these states are real or spurious
could be obtained more simply by examining the stability of their
energies against changes in the size of the square-integrable basis.
Resonant states produced by a square-integrable basis should
show stability \cite{LN}.

Just to give an example that the incorrect boundary
condition can incur spurious states,
here we show the case of $^8$Li. This nucleus is described
in a three-cluster $\alpha +t+n$ model with the basis containing a
number of different angular momenta \cite{Li8}. Expanding all
intercluster relative-motion functions in terms of square-integrable
functions, one gets, from the diagonalization
of the Hamiltonian, two $2^+$ states below the $\alpha +t+n$
threshold, at binding energies 4.2 MeV and 1.1 MeV (relative
to the three-cluster threshold), respectively. Experiment has
only produced one $2^+$ state in this region, the ground state.
The square-integrable basis is seemingly adequate, because we
are below the three-cluster threshold. However, the
1.1 MeV-energy state is above the $^7$Li+$n$ two-body
break-up threshold, which means that the correct boundary condition
has to contain a $^7$Li+$n$ scattering term. Supplementing
our wave function by such a term, the 1.1 MeV
$2^+$ state disappears immediately, showing that this state
was an artifact brought about by the incorrect boundary condition.

By this analogy, we suggest that the low-lying positive-parity
states of $^5$He could also disappear if the proper boundary
condition were taken into account. Be that as it may, we cannot
claim that this would disprove the existence of the states in question.
But certainly, before further efforts are spent on understanding the
nature of these states, the empirical evidence for their existence
should be reconsidered.

In conclusion, the reproduction of the negative quad\-ru\-pole
moment of $^6$Li and the low-lying positive-parity states
of $^5$He by Zheng {\it et al.} \cite{Zheng} in a shell model cannot
be regarded as well-founded because the aspects of few-body dynamics
underlying these effects are treated improperly in that model.

\mbox{}

This work was supported by NSF Grant Nos. PHY92-53505 and
PHY94-03666, and by the OTKA Grant No. 3010. We wish to thank
Dr. A.~T. Kruppa for useful discussions.

\begin{figure}
\caption{The quadrupole moment of the ground state of $^7$Li
as a function of its $\alpha +t$ separation energy in an
$\alpha +t$ cluster model. The experimental separation energy
is 2.47 MeV, and the quadrupole moment is $-$4.06 $e$$\cdot$fm$^2$
\protect\cite{Ajzenberg}.}
\end{figure}


\begin{references}
\bibitem{note} Note that, although the model of Ref.\
\protect\cite{Zheng} does not contain threshold energies and relative
motions explicitly, a projection of the wave
function onto an $\alpha+d$ model space should involve an $\alpha$--$d$
relative-motion function, and the fall-off of the tail of this function
should bear the $\alpha +d$ separation energy.
\end{references}
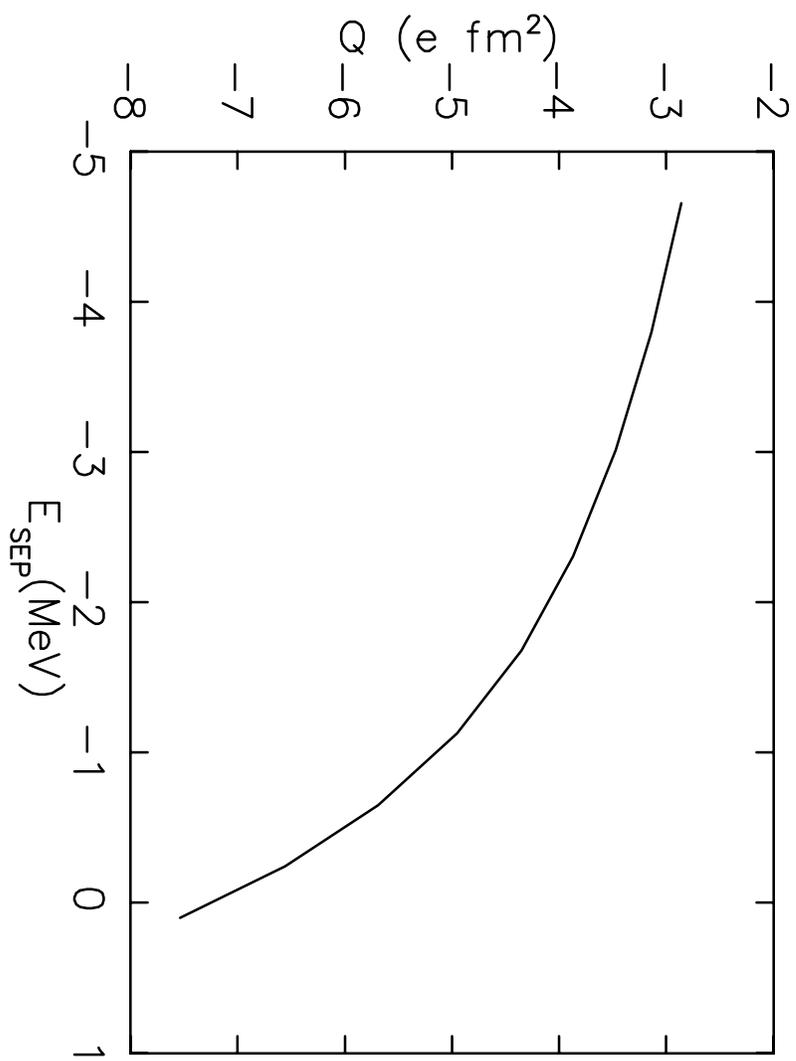
\end{document}